# TUNING AND FIELD SENSITIVITY OF Π-MODE STANDING WAVE LINACS FOR THE NLC [†]

R.M. Jones, R.H. Miller, J.W. Wang, and P.B. Wilson

†Stanford Linear Accelerator Center,
2575 Sand Hill Road, Menlo Park, CA, 94025

Abstract

In order to mitigate the effects of electrical breakdown (which have been found to occur in SLAC X-band traveling wave structures) standing wave structures are being considered for the NLC linac. At SLAC, structures consisting of 15 cells operating in the π accelerating mode are being tested for their electrical breakdown characteristics. In this paper the tuning requirement on the cavities is elucidated by utilizing a circuit model of the structure. The sensitivity of the field to both random and systematic errors is also discussed.
.

*Paper presented at the 2002 8th European Particle Accelerator Conference (EPAC 2002)*
*Paris, France,*
*June 3rd -June 7th, 2002*

This work is supported by Department of Energy grant number DE-AC03-76SF00515[†]

# TUNING AND FIELD SENSITIVITY OF Π-MODE STANDING WAVE LINACS FOR THE NLC†

R.M. Jones, R.H. Miller, J.W. Wang, and P.B. Wilson, SLAC, Stanford, CA94309, USA

*Abstract*

In order to mitigate the effects of electrical breakdown (which have been found to occur in SLAC X-band traveling wave structures) standing wave structures are being considered for the NLC linac. At SLAC, structures consisting of 15 cells operating in the π accelerating mode are being tested for their electrical breakdown characteristics. In this paper the tuning requirement on the cavities is elucidated by utilizing a circuit model of the structure. The sensitivity of the field to both random and systematic errors is also discussed.

## 1. INTRODUCTION

In order to accelerate multiple bunches of electron (and positron) beams up to a centre of mass of 500GeV, and later up to 1.0TeV or more, travelling wave (R)DDS accelerator structures have been designed and tested at SLAC and KEK [1]. However these structures have been found to suffer from electrical breakdown [2]. Recent design improvements, incorporating a reduced field on the surface of accelerating cavities and a reduced group velocity of the accelerating mode have been made and these have been found to reduce the number of breakdown events substantially. However, SW (Standing Wave) structures can achieve the same electron beam energy gain at significantly reduced gradients compared to their travelling wave counterparts. Initial experiments have indicated that the number of breakdowns is vastly reduced in SW structures. However, in fabricating several tens of thousands of SW structures, as will be required for the NLC (Next Linear Collider), it important to have a knowledge of the tuning characteristics of these structures as well as the sensitivity of the field to errors. This is investigated in the sections 2 and 3.

## 2 CIRCUIT MODEL OF SW STRUCTURE

### *2.1 Lumped circuit model of coupled cavities*

In order to calculate the modal frequencies and the flatness of the field we utilize a lumped circuit model of the N+1 coupled cavities illustrated in Fig. 1. In the designs under consideration at SLAC we have incorporated 15 cells operating at a π mode accelerating frequency of 11.424GHz, driven with RF field through the centre cell.

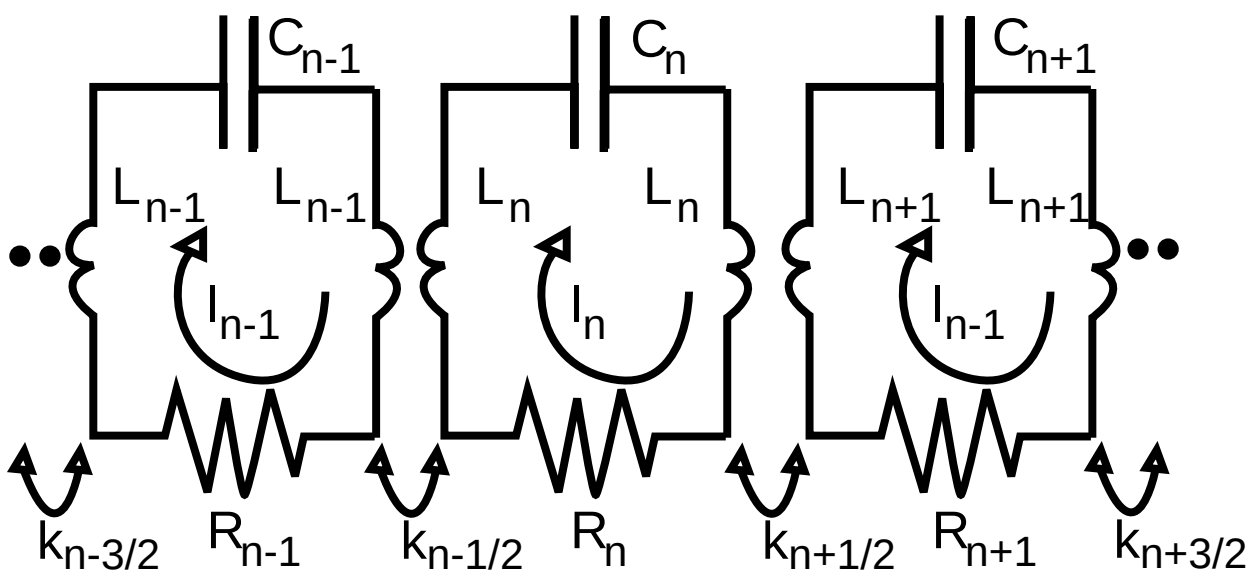

Figure 1: Circuit diagram of 3 cells in a SW accelerator.

The equations describing the zeroth (first cell), nth (all cell loops apart from first and last) and Nth (N=14 for a 15 cell cavity) loop, are given by:

$$(1-\frac{\omega_0^2}{\omega^2}+\frac{k_{-1/2}}{2})i_0+\frac{\omega_0'}{\omega_1'}\frac{k_{1/2}}{2}i_1=0 \qquad (2.1)$$

$$(1-\frac{\omega_n^2}{\omega^2})i_n+(\frac{k_{n-1/2}}{2}\frac{\omega_n'}{\omega_{n-1}'}i_{n-1}+\frac{k_{n+1/2}}{2}\frac{\omega_n'}{\omega_{n+1}'}i_{n+1})=0 \qquad (2.2)$$

$$(1-\frac{\omega_N^2}{\omega^2}+\frac{k_{N+1/2}}{2})i_N+\frac{\omega_N'}{\omega_{N-1}'}\frac{k_{N-1/2}}{2}i_{N-1}=0 \qquad (2.3)$$

where: $\omega_n^2=\omega_n'^2(1+j\frac{\omega}{\omega_n'}\frac{1}{Q_n})$, $\omega_n'^2=(2L_nC_n)^{-1}$, the quality factor is given by: $Q_n=2\omega_n'L_n/R_n$ and $k_{n+1/2}$ describes the n to n+1cell-to-cell coupling. We have used full cell boundary conditions: $i_{-1}=i_0$ and $i_{N+1}=i_N$.

In the test structures being built at SLAC $k_{n-1/2}=k_{n+1/2}$ =k, but in future structures the coupling will vary in order to detune and damp the dipole modes. We make the approximation that all reactive variation between cavities occurs in the inductance (i.e. the cell capacitances are fixed) and small frequency perturbations will be considered in this analysis. For identical cells eq. 2.2 reduces to:

$$i_n=(\text{cons.})\cos n\pi q/N \text{ and } \omega_q^2=\omega_c^2/(1+k\cos\pi q/N) \qquad (2.4)$$

where n (=0, N) is the cell number and q (=0, N) is the mode number. For non-identical cells, as is the case for all mistuned structures, we solve the non-linear eigensystem corresponding to eq 2.2. An iterative scheme converges very rapidly to the solution for the eigenvalues and eigenvectors when the Q values are large (as is true in practice where the Ohmic Q~8000).

Several structures have been fabricated and tuned, but here we present the results of measuring the field profile,

† Supported by Department of Energy grant number DE-AC03-76SF00515

via a bead pull technique, of SW20565 (a 15 cell structure in which the iris radius is 5.65mm). The field profile at each stage of the tuning process is shown in Fig. 2 beginning the tuning at A and ending at D. In order to calculate the frequency mistuning of the cells we wrote a Fortran computer code to minimize the "cost" function, defined as the difference between the sum of the squares of the circuit model eigenvectors and those of the experimentally determined field amplitude at each cell.

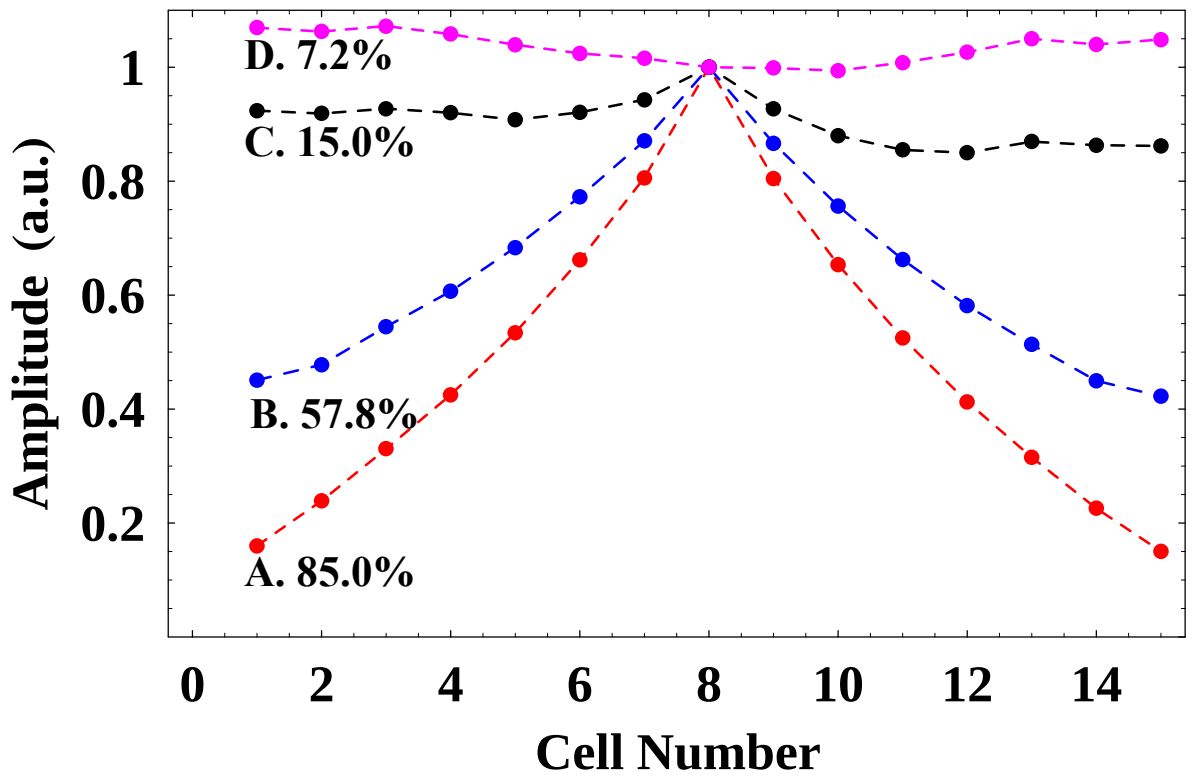

Figure 2: Experimental measurement of field amplitude at each stage of the tuning process. The maximum deviation of the field is also indicated on each curve.

This non-linear, 16-parameter (15 cells frequencies plus the accelerating mode frequency) minimization process converges after several hundred iterations and takes no more than a few minutes on a 400MHz PC and the residual in the cost function is no more than $10^{-9}$ (a.u.).

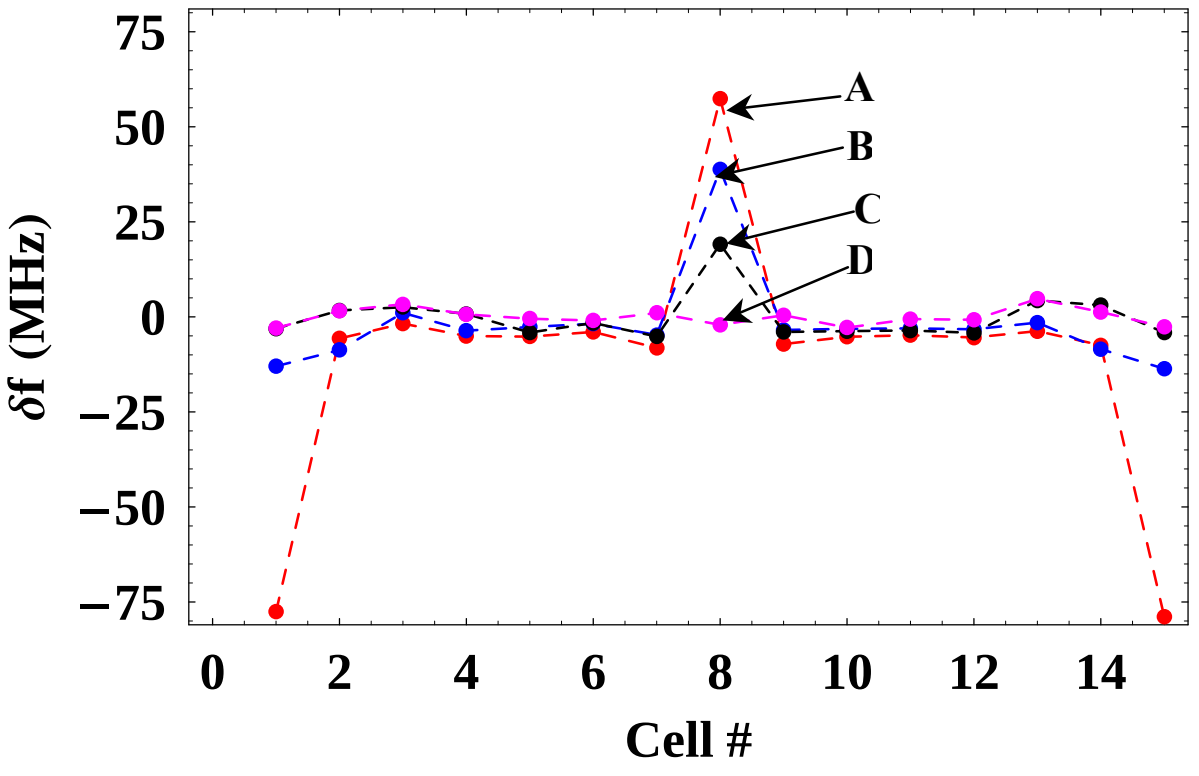

Figure 3: Cell frequencies at each stage of tuning process calculated from the eigenvalues of circuit.

The resulting differences in the cell frequencies from the perfectly flat, tuned condition, are shown in Fig 3. To begin with (curve A), the final cells are mistuned by 75MHz and the central (coupler cell) by more the 60MHz. However, after several anneals and tune-ups the final mistuning is no more than 4MHz. and the resulting field has an RMS deviation from perfectly flat of 7.2%. The field has a minimum in the coupler cell (the middle cell); this is desirable as electrical breakdown has been found to be concentrated in this region [2].

## *2.2 $S_{11}$ from an impedance model*

Viewed from the coupler waveguide, it can be shown that the reflection coefficient is given by:

$$S_{11} = (\beta - Y_c)/(\beta + Y_c) \qquad (2.5)$$

where $Y_c$ is the admittance of the circuit viewed from the cavity circuit and $\beta$ is the coupling coefficient defined as the ratio of the power coupled into the cavity to the power dissipated in the walls of the coupler. The admittance $Y_c$ is evaluated by taking the circuit of Fig. 1 and feeding current into the middle cell and calculating the voltage dropped across the equivalent reactive element. As we match into the cavity, at resonance we have:

$$S_{11} = (\beta - 1)/(\beta + 1) \qquad (2.6)$$

The coupling coefficient $\beta$ is calculated with Omega3P [3] and from this we are able to determine $S_{11}$ at resonance. The resonances of 2 further modes were measured and in Fig 4 the overall frequency response is compared with that predicted by the circuit model. The agreement is seen to be very good. As the SW structure is driven from the centre cell, then only odd modes are excited. This is confirmed in the driven-mode impedance model. Also shown in the figure are the eigenmodes of the closed system (source-less) and these line-up with the resonances of the driven response curves.

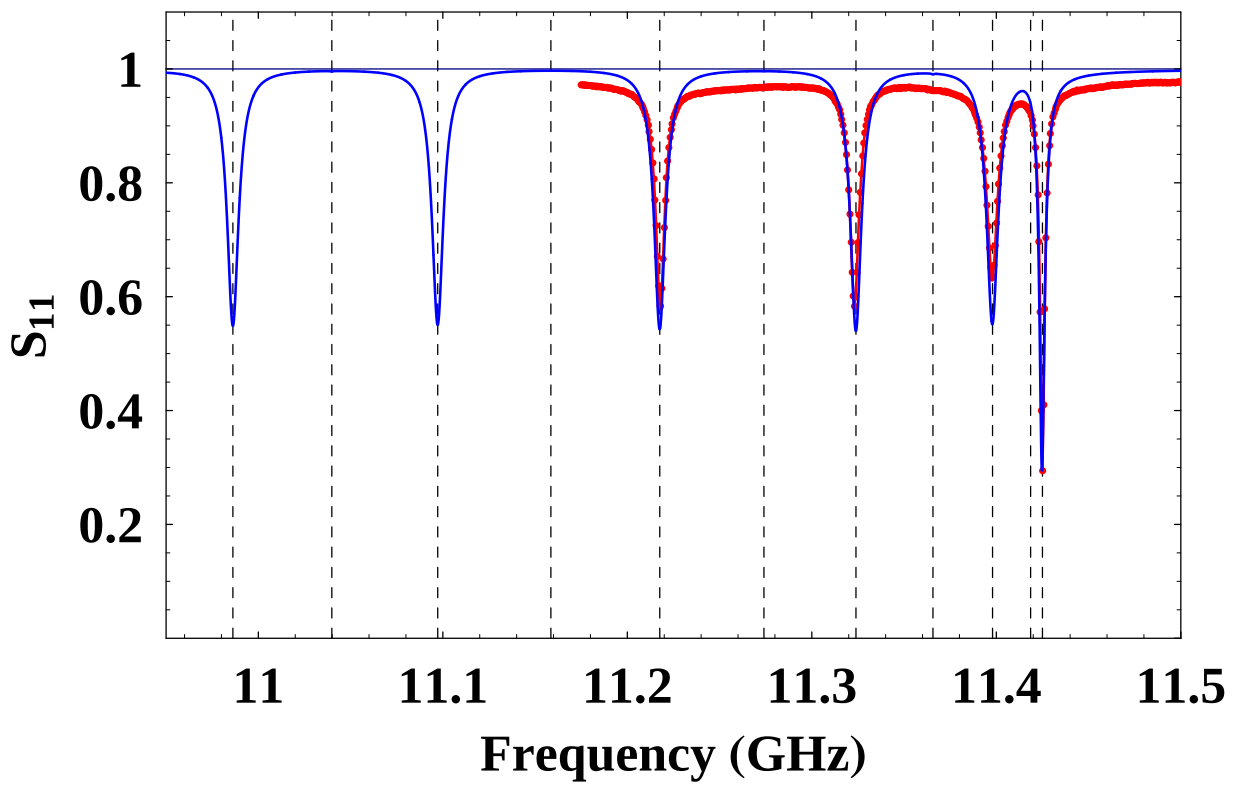

Figure 4: $S_{11}$ measured from centre port of 15-cell SW $\pi$ structure. The points (shown in red) are experimentally determined data and the line (blue) is obtained from the impedance model. Also shown, with dashed vertical lines, are 12 of the 15 eigenfrequencies.

# 3 FABRICATION ERRORS AND THEIR INFLUENCE ON FIELD FLATNESS

## *3.1 Systematic frequency errors*

In fabricating the structures a frequency error ($\delta f$) that is repeatable from cell-to-cell is classified as a systematic error. This error can be fixed by either retuning all the cavities to the correct resonance frequency or by placing a limited number of tuners within the accelerator and tuning the resonance frequency back to accelerating frequency. We have investigated several tuning schemes, but here we report on tuning the first and last cells together with the middle cell. The first and last cells are tuned by: $-\delta n \delta f/2$ (where $\delta n$ is number of cells present between tuners) and the middle cell is tuned by $\delta n \delta f$. This results in the accelerating frequency being recovered to within 100KHz. The resulting on-axis field profiles for several systematic errors are shown in Fig 5.

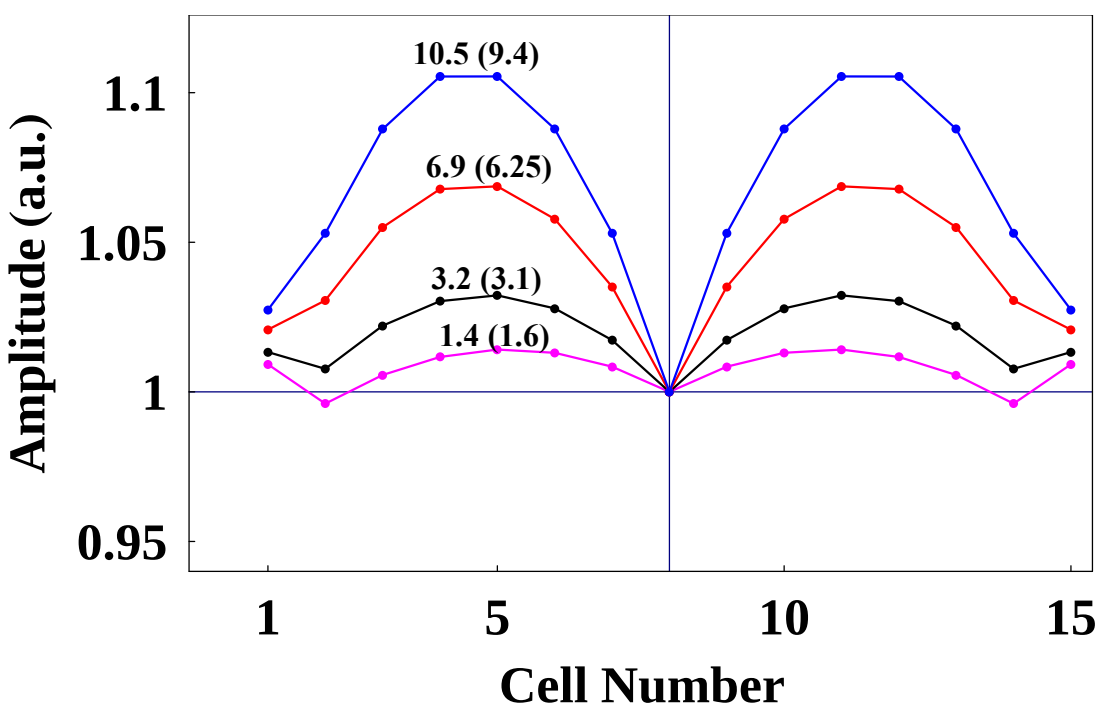

Figure 5: Standing wave structure with a systematic error of 0.5MHz (lowermost curve, magenta), 1, 2, and 3 MHz (uppermost curve, blue) included in the simulation. The maximum deviation of the field from perfectly flat is included on each of the curves and in parentheses an approximate result for the maximum deviation obtained from the analytic formula

An analytical formula for the maximum amplitude deviation along the complete structure can derived from the dispersion equation by considering the frequency shift in terms of the phase induced in the structure:

$$\delta f = \frac{\Delta f (\phi - \pi)^2}{4} \tag{3.1}$$

where $\phi$ is the phase advance per period along a structure with systematic frequency errors and $\Delta f \ (= f_\pi - f_0)$ is the bandwidth of the structure. The phase difference is then given by:

$$\delta\phi = \phi - \pi = 2\sqrt{\frac{\delta f}{\Delta f}} \tag{3.2}$$

The difference between the perfectly tuned and the structure with a field droop is now given by:

$$\frac{\Delta A}{A} = 1 - \cos(\delta n \Delta\phi / 2) = \frac{\delta n^2}{2} \frac{\delta f}{\Delta f} \tag{3.3}$$

Applying this approximate formula to the field in Fig 5 for a 1 MHz error gives a 3.1% field droop compared to 3.2% obtained from the exact calculation and thus this analytical result is certainly accurate enough to be used as a design tool to estimate how large the field droop is liable to be in a structure with limited tuners. The accuracy of this formula, of course improves as the systematic frequency error is reduced and as the number of tuners is increased.

### *3.2 Random frequency errors*

Random errors are introduced into each of the cells of the 15-cell structure and the field profile is recorded. In order to tune-up the field back to its original perfectly flat initial state we allow only three tuners to be used. These tuners are situated in the first, last and middle cavities. We seek to minimise the "cost" parameter, defined as the sum of the squares of the differences of the field from perfectly flat. And to this end, a computer program was written to automatically minimize the cost parameter. Introducing frequency errors which are uniformly distributed with an RMS error of 3MHz results in field distribution which deviates from perfectly flat with an RMS of 4%. This result was obtained through 50 simulations and is the mean deviation of the RMS of the deviation of the field from unity.

A typical result of this automated field tuning process is shown in Fig. 6. The field prior to tuning has an RMS deviation from unity of 17.8% and after tuning it has an RMS deviation of 1.6%. The frequency detuning required of the cells is shown inset to the figure and it is seen to be 8MHz for the centre cell and 2MHz for the end cell. These small frequency tunings are readily achievable in practise with small tuning screws attached to the cavities.

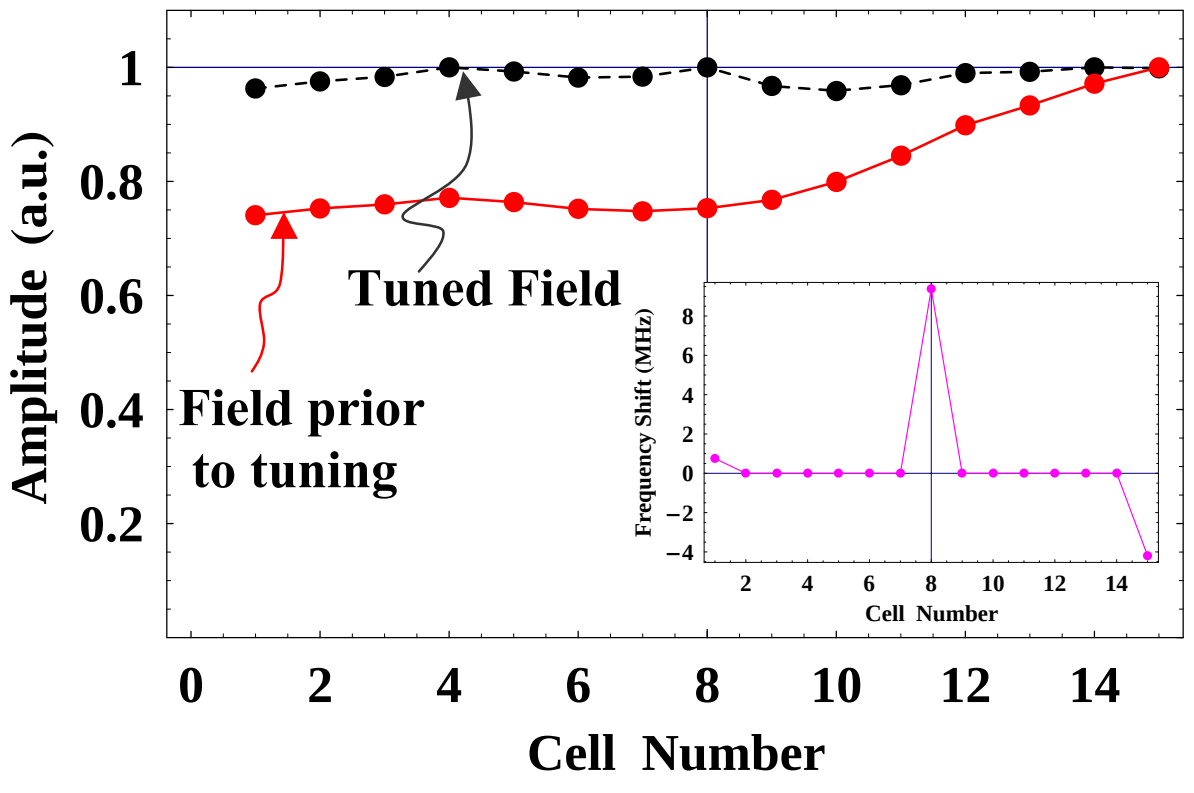

Figure 6: The amplitude of accelerating field under the influence of random errors with an RMS of 3MHz is illustrated. The first, last and middle cells are tuned in order to flatten out the 17.8% RMS deviation of the field from unity. Tuning 3 cells results in an RMS deviation from a perfectly flat field of 1.6%. Shown inset is the frequency shift of the cells required to flatten the field

## 4. REFERENCES

[1] J.W. Wang et al, LINAC2000, also SLAC-PUB 8583
[2] C. Adolphsen et al, PAC2001, also SLAC-PUB 8901
[3] Y. Sun., et al, ACES Conf., 2002, Monterey.